\documentstyle[12pt,aaspp4]{article}
\begin{document}
\def\msun{\hbox{${\cal{M}}_{\odot}$}}
\def\massA{\hbox{${\cal{M}}_A$}}
\def\massB{\hbox{${\cal{M}}_B$}}
\def\massAB{\hbox{${\cal{M}}_{A+B}$}}
% 21 Apr 2014, v1, raw text and tables only
% 03 Aug 2016, v2, text, tables and figures
% 08 Aug 2016, v3, document tightened up
%
% 14 Aug 2016 comments from Andrei Tokovinin
%
% 01 Jan 2017, v4, following AST2016 paper and including measures
% from Tokovinin et al. 2017. Orbits to GKI 3 and BLA 9 removed.
% First orbit for MTG 2 and masses calculated. Relevant figures
% redone.
%
% 02 Feb 2017 comments from John Subasavage
% 04 Mar 2017 comments from Jen Winters
% 10 Apr 2017 comments from Bill Hartkopf
% 13 Apr 2017 UCAC5 proper motion replace those from UCAC4 in paper.
% 11 May 2017 comments from Todd, v5
% 02 Jun 2017 comments from Todd
% 17 Jul 2017 MTG   2 orbit recalculated with a new point from
%             Andrei, fig, tab 3, tab 5 and text modified slightly
% 14 Sep 2017 MTG   4 orbit recalculated with a new point from
%             Andrei, fig, tab 3, tab 5 and text modified slightly
% 03 Oct 2017 Added correct reference from Bartlett et al. paper and
%             checked references in text to it to make sure they are
%             still correct.
% 06 Nov 2017 Tables 3, 4, 5 figures and text updated following 
%             additional measures in SOAR2016-7 paper draft.
% 07 Nov 2017 Changes following Editorial Board review.
% 20 Nov 2017 Final pre-submission corrections, v6
% 29 Jan 2018 Changes following Referee Report
% 17 Apr 2018 Changes following page proofs
% 
\title{Speckle Interferometry of Red Dwarf Stars}

\author{Brian D.\ Mason\altaffilmark{1}, William I.\ 
Hartkopf\altaffilmark{1,2}, Korie N.\ Miles\altaffilmark{3}}
\affil{U.S. Naval Observatory \\
3450 Massachusetts Avenue, NW, Washington, DC, 20392-5420 \\ 
Electronic mail: brian.d.mason@navy.mil}

\author{John P.\ Subasavage\altaffilmark{1,4}}
\affil{U.S. Naval Observatory \\
Flagstaff, AZ 86001 \\
jsubasavage@nofs.navy.mil}

\author{Deepak Raghavan\altaffilmark{1}}
\affil{Center for High Angular Resolution Astronomy \\
Georgia State University, P.O.\ Box 3969, Atlanta, GA 30302-3969 \\
raghavan@astro.gsu.edu}

\author{Todd J.\ Henry}
\affil{RECONS Institute \\
Chambersburg, PA 17201 \\
toddhenry28@gmail.com}

\altaffiltext{1}{Visiting Astronomer, Kitt Peak National Observatory and 
Cerro Tololo Inter-American Observatory, National Optical Astronomy 
Observatories, operated by Association of Universities for Research in 
Astronomy, Inc.\ under contract to the National Science Foundation.}

\altaffiltext{2}{Retired.}

\altaffiltext{3}{SEAP Intern.}

\altaffiltext{4}{Current address: The Aerospace Corporation, 2310 E.\ El
Segundo Boulevard, El Segundo, CA 90245, email: john.subasavage@aero.org.}

\begin{abstract}

%===========================================================================
We report high resolution optical speckle observations of 336 M 
dwarfs which result in 113 measurements of relative position of 80 
systems and 256 other stars with no indications of duplicity. These 
are the first measurements for two of the systems. We also present 
the earliest measures of relative position for 17 others. We include
orbits for six of the systems, two revised and four reported for the
first time. For one of the systems with a new orbit, G 161-7, we 
determine masses of 0.156$\pm$0.011 and 0.1175$\pm$0.0079\msun ~ for
the A and B components, respectively. All six of these new 
calculated orbits have short periods between five and thirty-eight 
years and hold the promise of deriving accurate masses in the near 
future. For many other pairs we can establish their nature as 
physical or chance alignment depending on their relative motion. Of 
the 80 systems, 32 have calculated orbits, 25 others are physical 
pairs, 4 are optical pairs and 19 are currently unknown.

% Red Dwarfs are common,
% but it's hard for them to have
% a little brother.

\end{abstract}


\section{Introduction}

%===========================================================================
Double stars are those stars which, seen through the telescope, 
present themselves as two points of light. Some of these are 
physically associated with each other and are true bona fide 
{\it binary stars}, while others are chance alignments. While these 
$``$optical doubles" may prove troublesome as stray light 
complicates both photometry and astrometry, they are astrophysically
inconsequential. The true binary nature of double stars can be 
detected through a variety of means, from wide systems found via 
common proper motion (CPM) to orbit pairs to the even closer systems, 
found through periodic variations in radial velocity or photometry. For 
generations, painstaking measurements of have been collected in catalogs
such as the Washington Double Star Catalog 
(hereafter WDS; Mason et al.\ 2001). The organization of significant
data sets of multiple stars is critical to understanding the 
outcomes of the star formation process as well as key to identifying
which systems promise fundamental astrophysical parameters, e.g., 
masses.

Red dwarfs, specifically, M dwarfs, are the most common stellar 
constituent of the Milky Way, accounting for three of every four 
stars (Henry et al.\ 2006). However, their binary fraction is quite 
low in comparison to other stars ($\sim$27\%; Winters et al.\ 2015).
The other end of the Main Sequence, the O stars, have a very 
high binary fraction (43/59/75\% for Runaway/Field/Cluster samples; 
Mason et al.\ 2009). Possible companions to an O star may include 
stars from the entire spectral sequence, while the only possible 
stellar companions to an M dwarf are lower mass M dwarfs, brown 
dwarfs or fainter evolved objects. Mass determinations of M Dwarfs
are poorly constrained\footnote{Although, 
thanks to work such as Benedict et al.\ (2016) it is getting better 
on the low-mass end.}, observations of M dwarfs, for binary 
detection, orbit determination, and eventual mass determination, are
of paramount importance. To improve the statistical basis for 
investigations of the nearest M dwarfs and to pinpoint systems 
worthy of detailed studies, in this paper, we report high resolution
optical speckle observations of 336 M dwarfs. We report 113 resolved
measurements of 80 systems, nineteen of these have their first 
measure reported here, although all but two of those have their 
first published measure elsewhere.

\section{Instrumentation and Calibration}

%===========================================================================
Observing runs for this program are provided in Table 1,
which includes the dates, telescopes and observers, a
subset of the authors on this the paper. The observing runs included many different projects since 
speckle interferometry is a fast observing technique with up to 20 
objects per hour observed and nightly totals of 120-220 stars 
depending on hours of dark time. Most data not specific to this M dwarf
program were Massive stars (Mason et al.\ 2009) or Exoplanet hosts
(Mason et al.\ 2011). Other data are presented in Appendix A. 
The instrument used for these observations was the USNO speckle 
interferometer, which is described in detail
in Mason et al.\ (2009, 2011). Briefly, the camera consists of two 
different microscope objectives giving different scales, interference 
filters of varying FWHM to allow fainter objects to be observed, Risley 
prisms which correct for atmospheric dispersion and finally a Gen IIIc ICCD 
capable of very short exposures necessary to take advantage of the 
$``$speckling" generated by atmospheric turbulence. Each observation 
represents the directed vector autocorrelation (Bagnuolo et al.\ 1992) of 
2000$+$ individual exposures, each $1-15msec$ long, depending on an 
object's brightness and the filter in use.  As the speckles are an 
atmospheric effect independent of the telescope, a larger telescope sees more 
turbulence cells and, therefore, more speckles. While a larger telescope 
can produce more correlations and a higher SNR it does not significantly 
change the magnitude limit. Brighter primary stars
with $V~<11.5$ were observed with a {\it Str\"{o}mgren y} filter (FWHM 
$25nm$ centered on $550nm$). Stars fainter than this were observed with 
a {\it Johnson V} filter (FWHM $70nm$ centered on $550nm$). The resolution 
limit with the 4m telescope employed in these observations is $30mas$; 
however, when the wider filter was used, the resolution capability 
is degraded to $50mas$ due to the greater atmospheric dispersion. The
field of view is 1\farcs8 centered on the target. The camera is 
capable of multiple observing modes, where wider pairs, if seen in 
the field, can be observed and measured using 2$\times$2 or 
4$\times$4 binning\footnote{Increasing the field-of-view to 
3\farcs6 or 7\farcs2 in the horizontal or vertical and even larger
by $cos\theta$ along diagonals.}. However, this is only when the 
companion is seen or known {\it a priori}. In terms of the search 
for new companions the field-of-view is characterized as 
1\farcs8$\times$1\farcs8.

For calibration, a double-slit mask was placed over the ``stove pipe"
of the KPNO Mayall Reflector, and a known single star was observed. 
This application of the well known experiment of Young allowed for 
the determination of scale without relying on binaries themselves to
determine calibration parameters. The slit-mask, at the start of the 
optical path, generates peaks based upon the the slit-separation and the
wavelength of observation. These peaks can be measured using the same 
methodology as a double star measure and, thus, generates a very precise
scale for the CCD. See McAlister et al.\ (1987) \S 4 and Figure 4 for further
details. Multiple observations through the 
slit mask yield an error in the position angle zero point of 0\fdg20
and a scale error of 0.357\%. These ``internal errors" are 
undoubtedly underestimates of the true errors of these observations.
While this produces excellent calibration for the Mayall Reflector, due
to small differences between it and the CTIO Blanco Reflector, the double
slit-mask could not be placed on the CTIO 4m $``$stove pipe".
Because this option was not available on the CTIO Blanco Reflector, 
a large number of well-known equatorial binaries with very 
accurate orbits were observed with both 
telescopes to allow for the determination of more realistic global
errors. Given the long time between some of these observations, 
wider pairs were observed with other telescopes that were slowly 
orbiting and well-characterized, as well as linear pairs, were observed. This 
process prevented excessive extrapolation when measuring the
scale of the observed field. 

Speckle Interferometry is a technique that is sensitive to 
changes in observing conditions, particularly coherence length 
($\rho_0$) and time ($\tau_0$). These typically manifest as a 
degradation of detection capability close to the telescope 
resolution limit or at larger magnitude differences between 
components. To ensure we reached our desired detection 
thresholds, a variety of systems with well-determined and 
characterized morphologies and magnitude differences were observed 
throughout each observing night. In all cases, results for these 
test systems indicated that our observing met or exceeded the desired 
separation and magnitude difference goals. Most, but not all, of the 
systems observed for characterizing errors or investigating detection space 
were presented in Mason et al.\ (2011). Others are presented in Appendix A 
below. Overall, our speckle observations are generally 
able to detect companions to M dwarfs from $30mas~<~\rho~<1\farcs8$ 
if the $\Delta$m$_{\rm v}~<~2$ for M dwarfs brighter than $V=11.5$. 
If fainter than this, the resolution of close pairs is degraded such
that the effectively searched region is $50mas~<~\rho~<1\farcs8$.
Some observations and measurements were obtained during times of 
compromised observing conditions. Non-detections made at this time are not 
considered definitive and are not tabulated below. 

\section{Results}

Table 2 lists the astrometric measurements (T, $\theta$, and 
$\rho$) of the observed red dwarf stars. The first two columns identify the 
system by providing the WDS designation (based on epoch-2000 
coordinates) and discovery designation. Columns three through five 
give the epoch of observation (expressed as a fractional Julian 
year), the position angle (in degrees), and the separation (in 
seconds of arc). Colons indicate measures with reduced accuracy due
to observing conditions. Note that the position angle has not been 
corrected for precession, and thus, is based on the equinox for the 
epoch of observation. The sixth column indicates the number of 
observations contained in the mean position. Columns seven and eight
list position angle and separation residuals (in degrees and 
arcseconds, respectively) to the orbit or rectilinear fit referenced
in Column nine. Finally, the last column is reserved for notes for 
these systems. 

While some published orbits may be premature and some linear 
determinations may reflect relative motion of an edge-on and/or 
long-period eccentric binary, these are nominally used to 
characterize each pair as physical and optical, respectively. Other 
pairs, as indicated in the notes to Table 2, are further classified 
as physical or optical based on the relative motion of the pair through
inspection of their double star measures compared with the proper motion.
The proper motion of these M dwarfs are typically large, therefore double star
measures at approximately the same position over a time base of many
years establishes the pair as physical through common proper motion. This
assessment depends on the magnitude of the proper motion, the change in 
relative position, and the time between observations. This sort of analysis 
cannot be made for unconfirmed pairs.

For twenty-one of the pairs in Table 2 this represents the earliest
measure. While the data presented in Table 2 has not been published
before, their results had been shared with collaborators (Hartkopf et al.\
2012, Tokovinin et al.\ 2010, 2014, 2015, 2016, 2018, 2019). In addition,
independent initiatives of others (Benedict et al.\ 2016, Henry et al.\
1999, Horch et al.\ 2010, 2011, 2012, Janson et al.\ 2012, 2014a, 2014b,
Jodar et al.\ 2013, Riedel et al.\ 2014, Ward-Duong et al.\ 2015, Winters
et al.\ 2011, 2017) has further enhanced the capability to assess the
physicality of these pairs and have enabled many of the orbits and linear 
solutions presented below. 

Overall, 336 M dwarfs were observed. From these observations, we
completed 113 measures of position angle and separation for 80 different 
pairs.

\section{Analysis of Resolved Doubles}

\subsection{New Orbital Solutions}

All orbits were computed using the ``grid search" routine described
in Hartkopf et al.\ (1989); weights are applied based on the methods
described by Hartkopf et al.\ (2001a). Briefly, weights of the 
individual observations are evaluated based on the separation relative to the 
resolution capability of the telescope (larger telescopes produce more 
accurate data), the method of observation (e.g., micrometry, photography, 
interferometry, etc.), whether the published measure is a mean of multiple 
nights, and if the measurer made any notes regarding the quality of the 
observation. Elements for these systems 
are given in Table 3, where columns (1), (2) and (3) give the WDS 
and discovery designations, followed by an alternate designation; 
columns (4) -- (10) list the seven Campbell elements: $P$ (period, 
in years), $a$ (semi-major axis, in arcseconds), $i$ (inclination, 
in degrees), $\Omega$ (longitude of node, equinox 2000.0, in 
degrees), $T_0$ (epoch of periastron passage, in fractional 
Julian year), $e$ (eccentricity), and $\omega$ (longitude of 
periastron, in degrees). Formal errors are listed with each element.
Columns (11) and (12) provide the orbit grade (see Hartkopf et al.\ 
2001a) and the reference for a previous orbit determination, if one 
exists. Orbit grades are on a $1-5$ scale. In the case of the orbits
presented here, a grade of $3$ indicates the orbit is $``$reliable,"
$4$ is $``$preliminary" and $``$5" is $``$indeterminate." In all 
cases here, the numbers are indicative of the small number of 
observations and incomplete phase coverage.

Figure 1 illustrates the new orbital solutions for the six systems whose
orbits are presented here, plotted together 
with all published data in the WDS database as well as the 
previously unpublished data from Table 2. In each of these plots, 
micrometric observations are indicated by plus signs, and 
photographic measures by asterisks; Hipparcos measures are indicated
by the letter `H', conventional CCD measures by triangles, 
interferometric measures by filled circles, and the new measures 
presented in Table 2 are indicated with stars. ``$O-C$" lines 
connect each measure to its predicted position along the new orbit 
(shown as a thick solid line). Dashed ``$O-C$" lines indicate 
measures given zero weight in the final solution. A dot-dash line 
indicates the line of nodes, and a curved arrow in the lower right 
corner of each figure indicates the direction of orbital motion. The
scale, in arcseconds, is indicated on the left and bottom of each
plot. Finally, if there is a previously published orbit it is shown 
as a dashed ellipse. The sources of those orbits are listed in the 
final column of Table 3.

\begin{figure}[p]
\begin{center}
{\epsfxsize 2.5in \epsffile{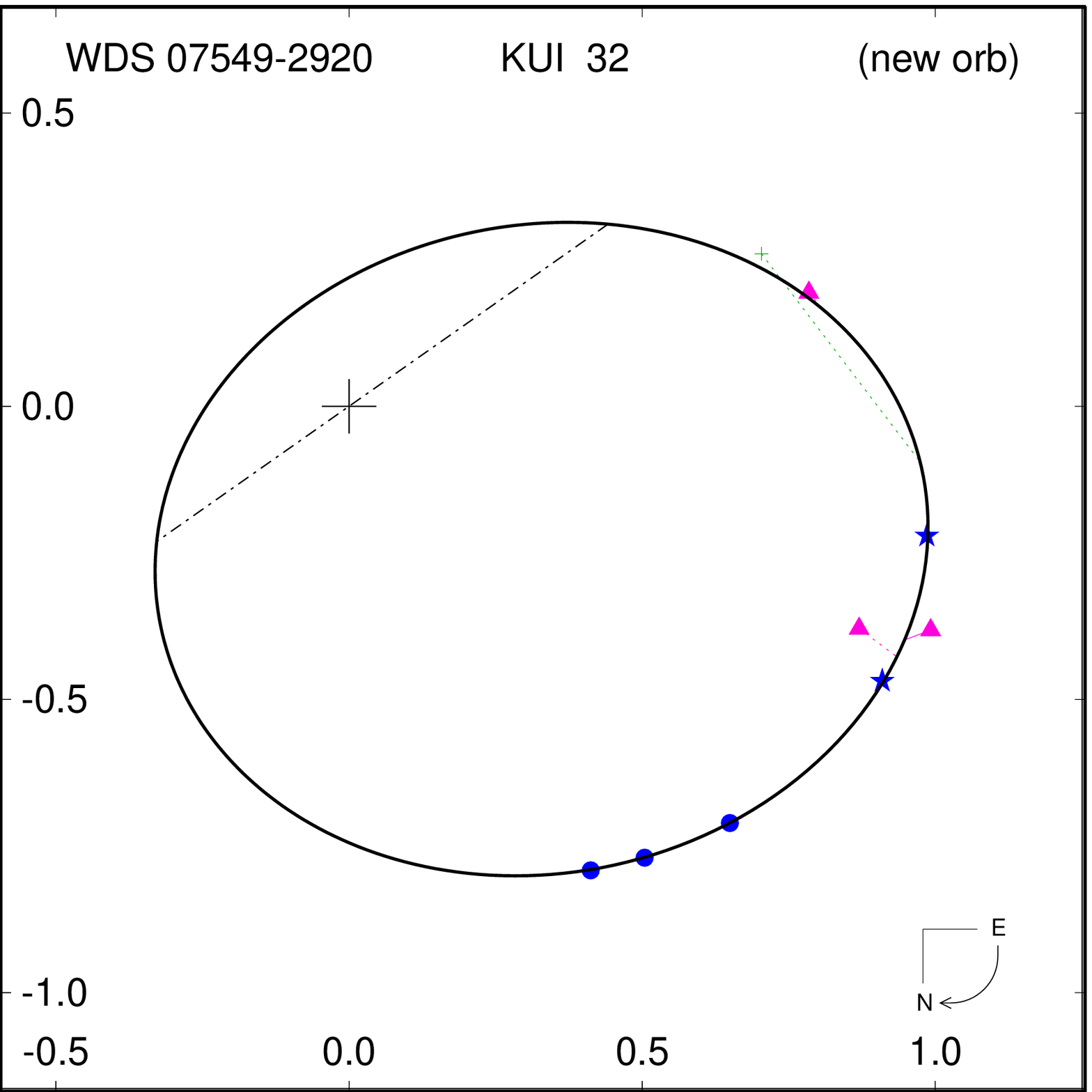} \epsfxsize 2.5in \epsffile{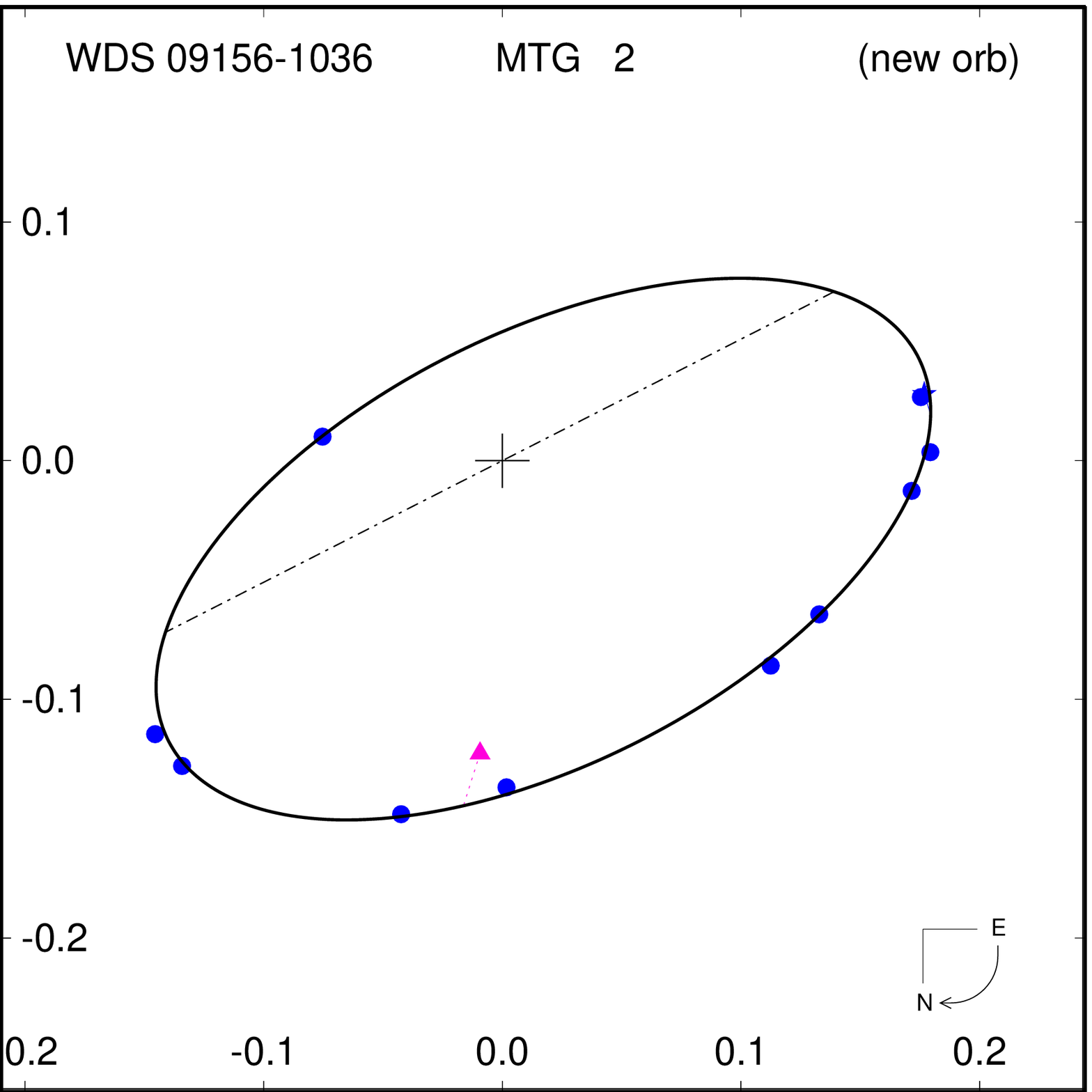}}
{\epsfxsize 2.5in \epsffile{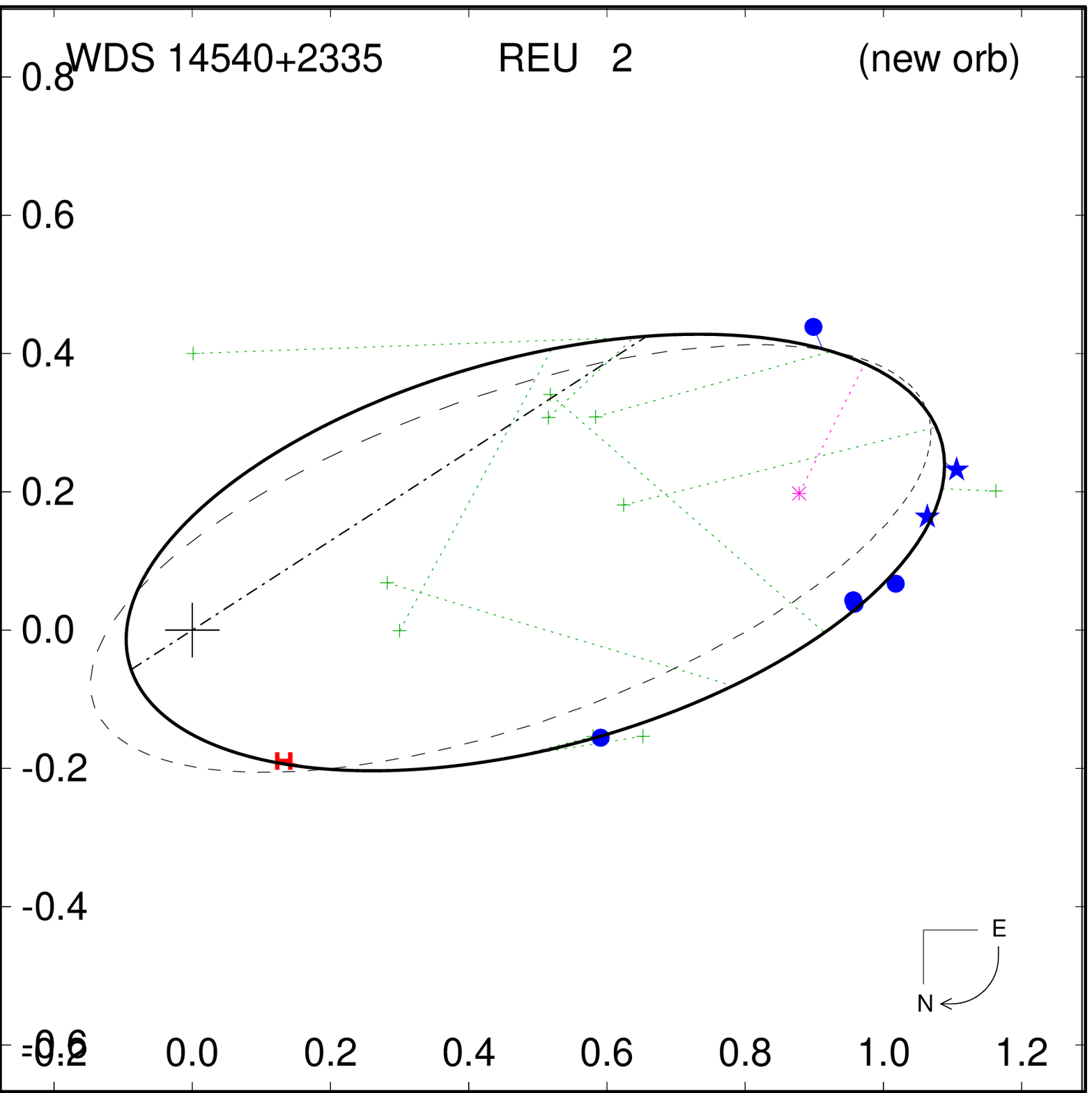} \epsfxsize 2.5in \epsffile{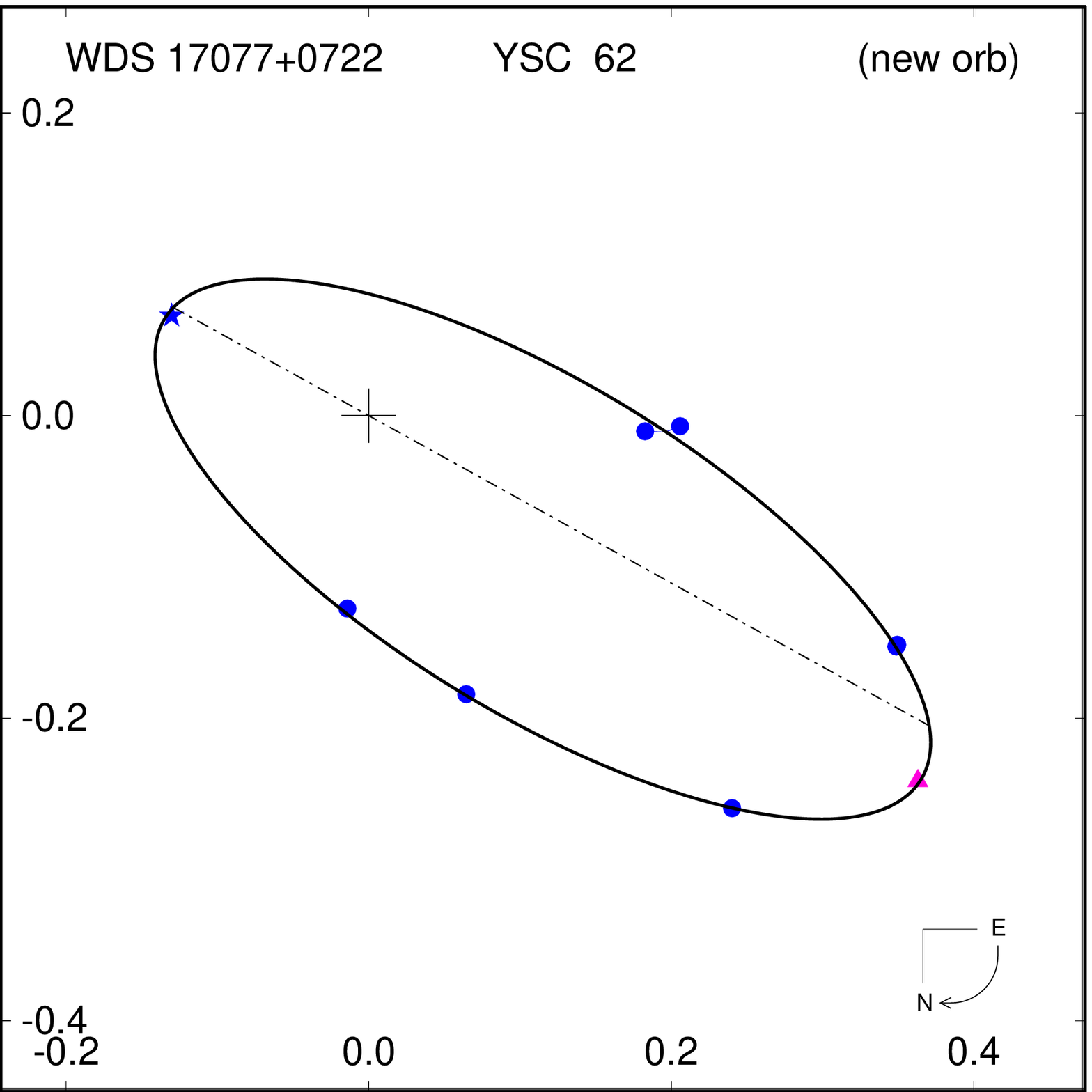}}
{\epsfxsize 2.5in \epsffile{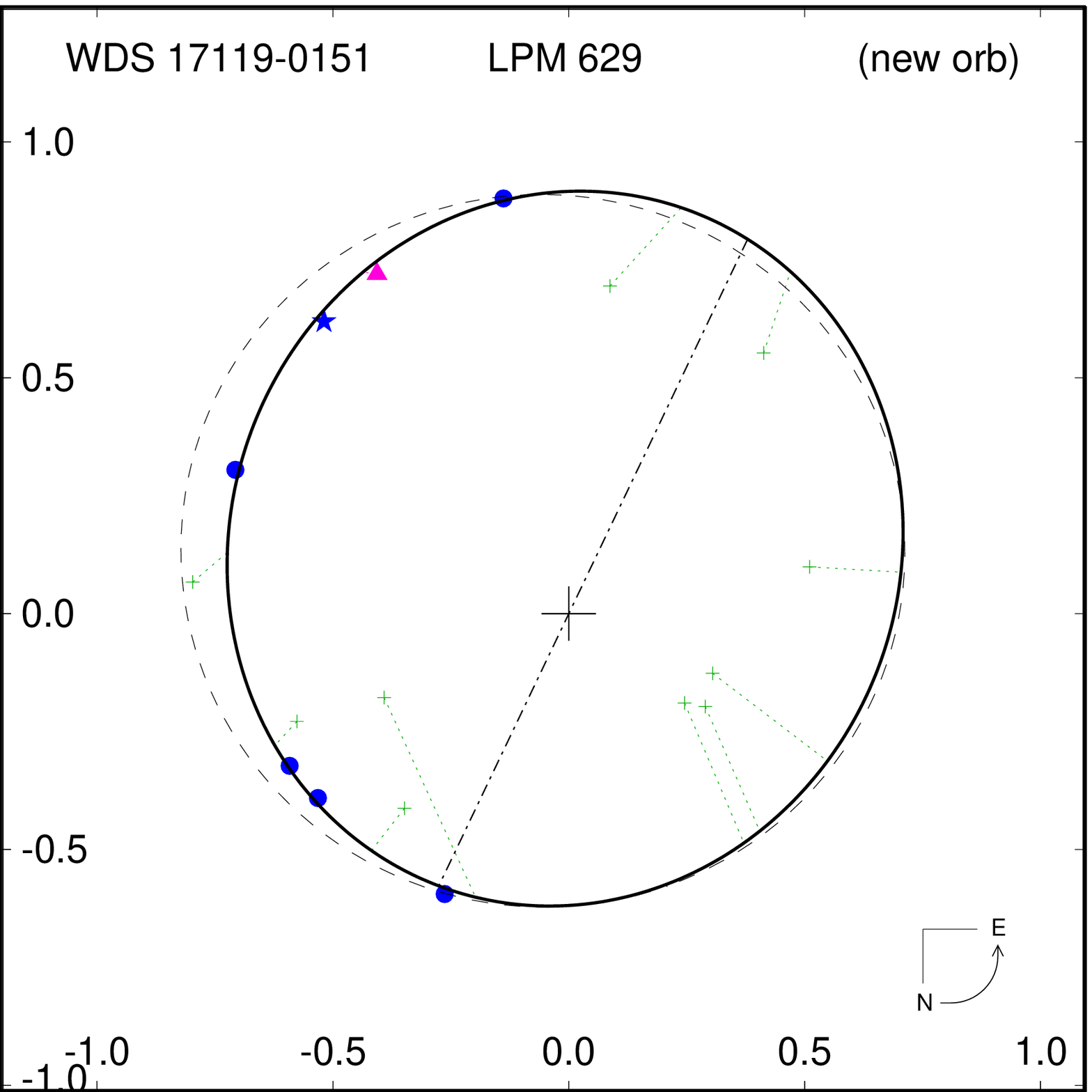} \epsfxsize 2.5in \epsffile{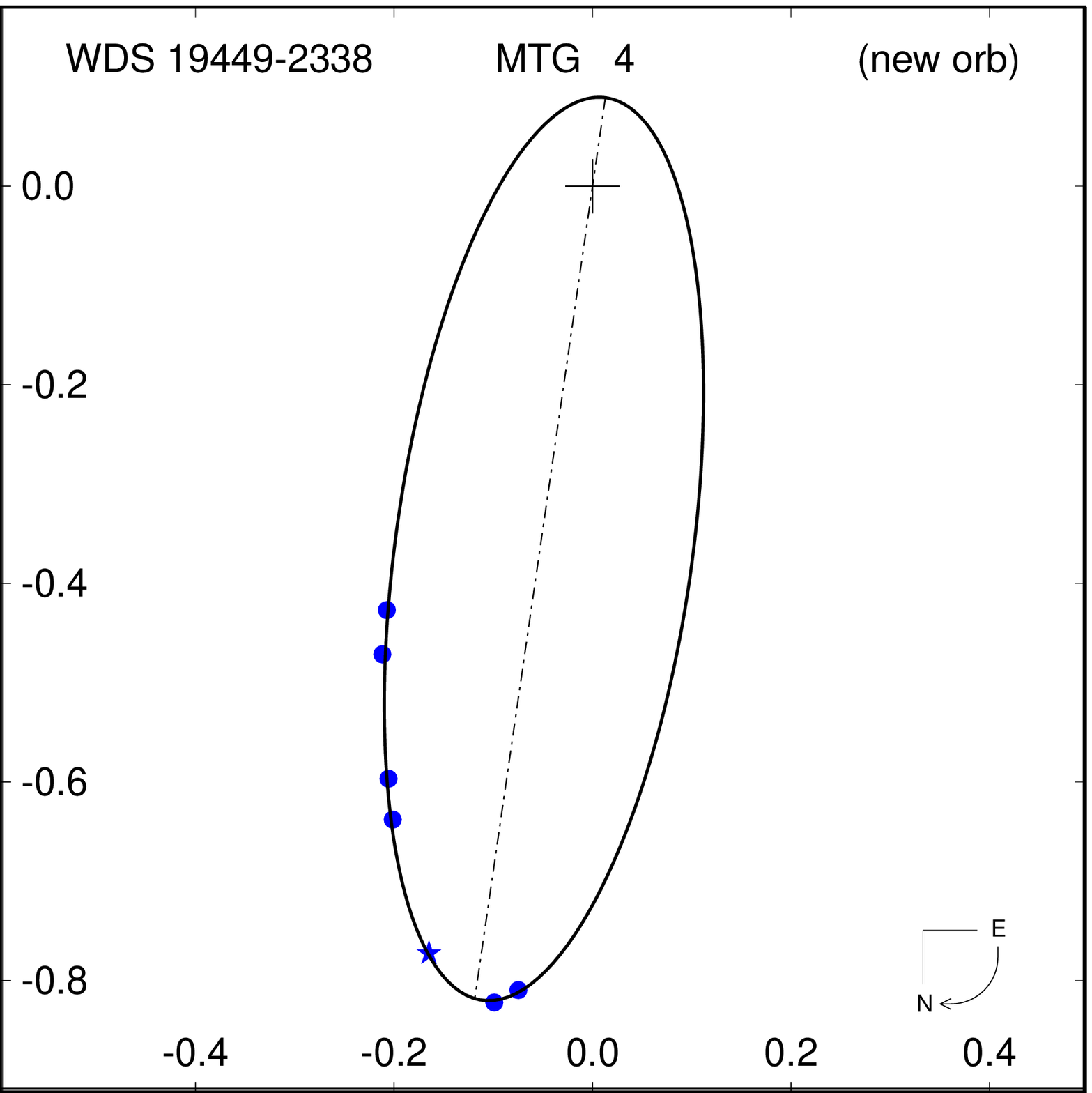}}
\end{center}
\caption{\small New orbits for the systems listed in Table 3 and all
data in the WDS database and Table 2. Micrometric observations are 
indicated by plus signs, and photographic measures by asterisks; 
Hipparcos measures are indicated by the letter `H', conventional CCD
measures by triangles, interferometric measures by filled circles, 
and the new measures presented in Table 2 are indicated with stars. 
``$O-C$" lines connect each measure to its predicted position along 
the new orbit (shown as a thick solid line). Dashed ``$O-C$" lines 
indicate measures given zero weight in the final solution. A 
dot-dash line indicates the line of nodes, and a curved arrow in the
lower right corner of each figure indicates the direction of orbital
motion. The scale, in arcseconds, is indicated on the left and 
bottom of each plot. Finally, if there is a previously published 
orbit, it is shown as a dashed ellipse.}
\end{figure}

The orbital periods of all six pairs (three of which have very high 
eccentricities; $> 0.7$) are all quite short, from 5$-$38y, and have 
small semi-major axes (0\farcs2-0\farcs9). The potential for 
improvement of the orbits and precise mass determinations for these 
pairs, all with large parallaxes, is excellent, especially for 
precise high angular resolution work with large aperture 
instruments. The errors of some of the earlier micrometry measures 
are quite high (e.g.\ WDS14540$+$2335), and are given quite low 
weight in the orbit. However, these historic observations can be 
quite helpful, especially in determining the orbital period. 
The most interesting of these six pairs is discussed in detail 
below while the remaining five are noted in \S6.

\subsubsection{G 161-7}

The M dwarf star G 161-7 (alternatively known as LHS 6167 or NLTT 
21329) was first resolved as a double with adaptive optics by 
Montagnier et al.\ (2006), who resolved the pair on two occasions. 
If the resolved optical companion of G 161-7 were simply a chance
alignment with small proper motion, then the high proper motion of G
161-7 would result in a relative shift of 1\farcs6 between the two 
components. However, the companion continues to stay quite close,
making this a very likely physical pair. While maintaining their 
proximity, large changes in the position angle of the companion 
demonstrated that the orbital period was short. Observed by this 
effort in 2010 (Table 2) the measures were also supplemented by 
Janson et al.\ (2014a) who observed it with $``$lucky imaging" and 
were able to split the pair as well as determine a mass ratio: 
0.57$\pm$0.05. Lately, it has been regularly observed by the 
SOAR-Speckle program (Tokovinin et al.\ 2015, 2016, 2018, 2019). 

Barlett et al.\ (2017) measured the parallax ($103.33\pm1.00mas$)
to this nearby pair and also made an estimate of $\sim$4y for the 
orbital period. Taking the available relative astrometry an orbital
solution with a period just over 5y quickly converged (see Table 3 and
Figure 1). With the parallax a mass sum is 0.273$\pm$0.018\msun ~ is
determined and with the mass ratio individual masses of 0.156$\pm$0.011
and 0.1175$\pm$0.0079\msun ~ are determined for A and B, respectively.
While Gaia parallax should be quite precise for this pair, the errors
of the orbit, already under 2\%, can be improved with the accumulation 
of more data filling in unobserved regions of the orbit. With this, the
orbital elements and, hence, the mass errors will improve. This pair is
the best example of what we hope this effort will ultimately achieve.
 
\subsection{New Linear Solutions}

Inspection of all observed pairs with either a 30$^{\circ}$ change 
in their relative position angles or a 30\% change in separations 
since the first observation cataloged in the WDS revealed six pairs
whose motion seemed linear. These apparent linear relative motions 
suggest that these pairs are either composed of physically unrelated
stars or have very long orbital periods. Linear elements to these 
doubles are given in Table 4, where Columns one and two give the WDS
and discoverer designations and Columns three to nine list the seven
linear elements: x$_{0}$ (zero point in x, in arcseconds), a$_{x}$ 
(slope in x, in $''$/yr), y$_{0}$ (zero point in y, in arcseconds), 
a$_{y}$ (slope in y, in $''$/yr), T$_{0}$ (time of closest apparent 
separation, in years), $\rho_{0}$ (closest  apparent separation, in 
arcseconds), and $\theta_{0}$ (position angle at T$_{0}$, in 
degrees). See Hartkopf \& Mason (2015) for a description of all 
terms. 

Figure 2 illustrates these new linear solutions, plotted together 
with all published data in the WDS database, as well as the 
previously unpublished data from Table 2. Symbols are the same as in 
Figure 1. In the case of linear plots, the dashed line indicates the 
time of closest apparent separation. As in Figure 1, the direction of 
motion is indicated at lower right of each figure. As the plots and solutions
are all relative, the proper motion ($\mu$) difference is assumed to
be zero.

\begin{figure}[p]
~\vskip -1.8in
\begin{center}
{\epsfxsize 2.8in \epsffile{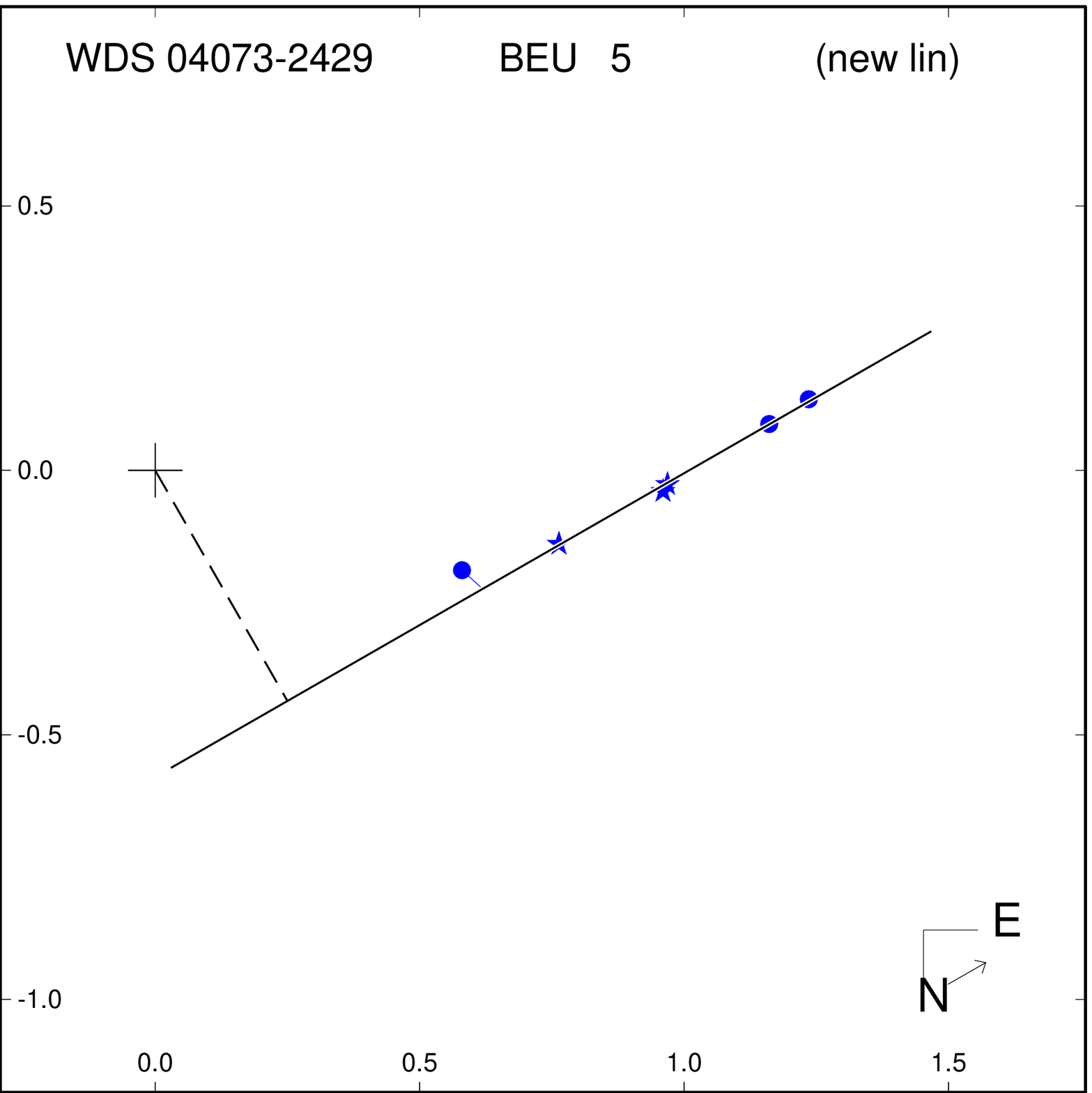} \epsfxsize 2.8in \epsffile{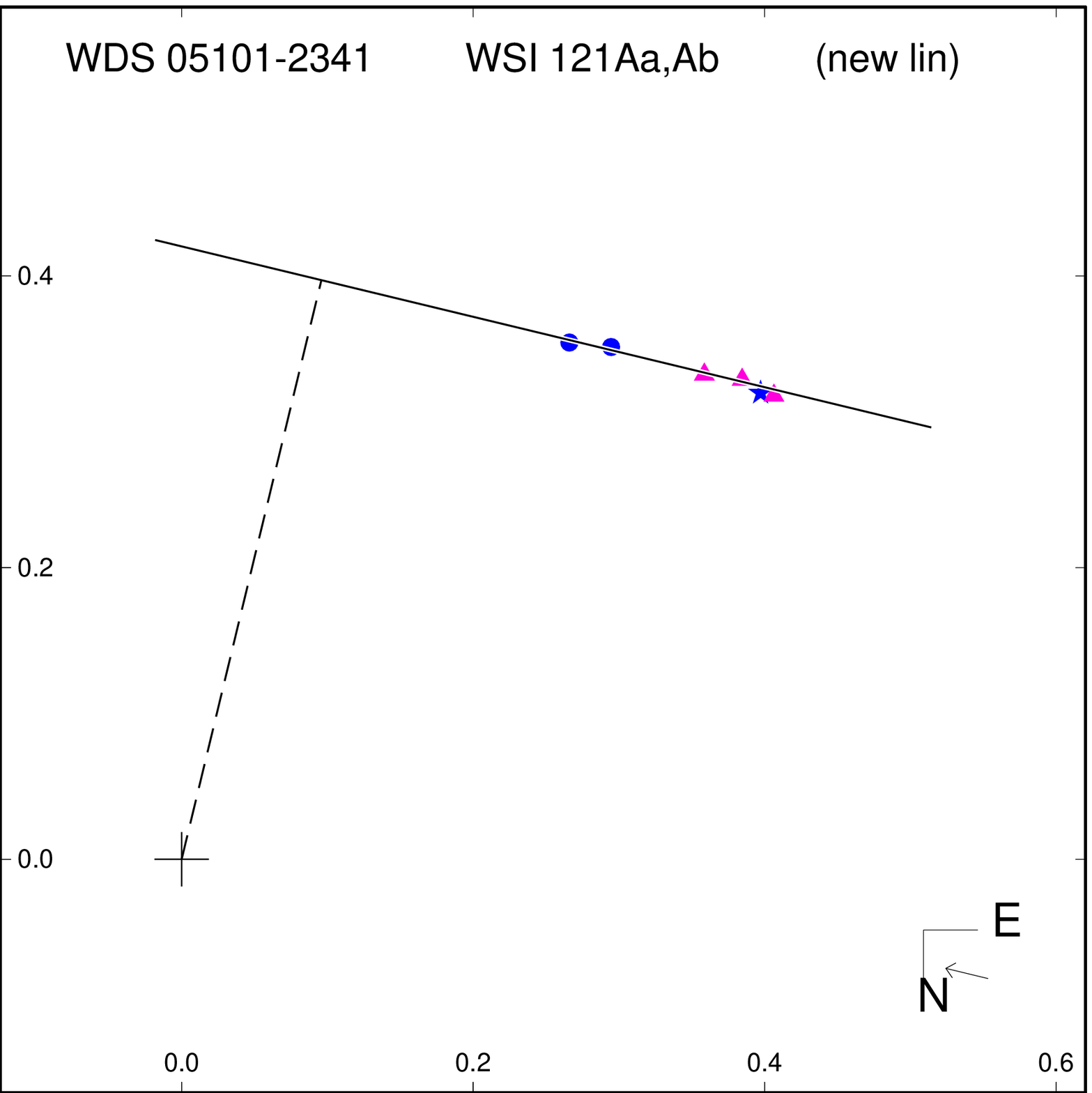}}
\vskip 0.05in
{\epsfxsize 2.8in \epsffile{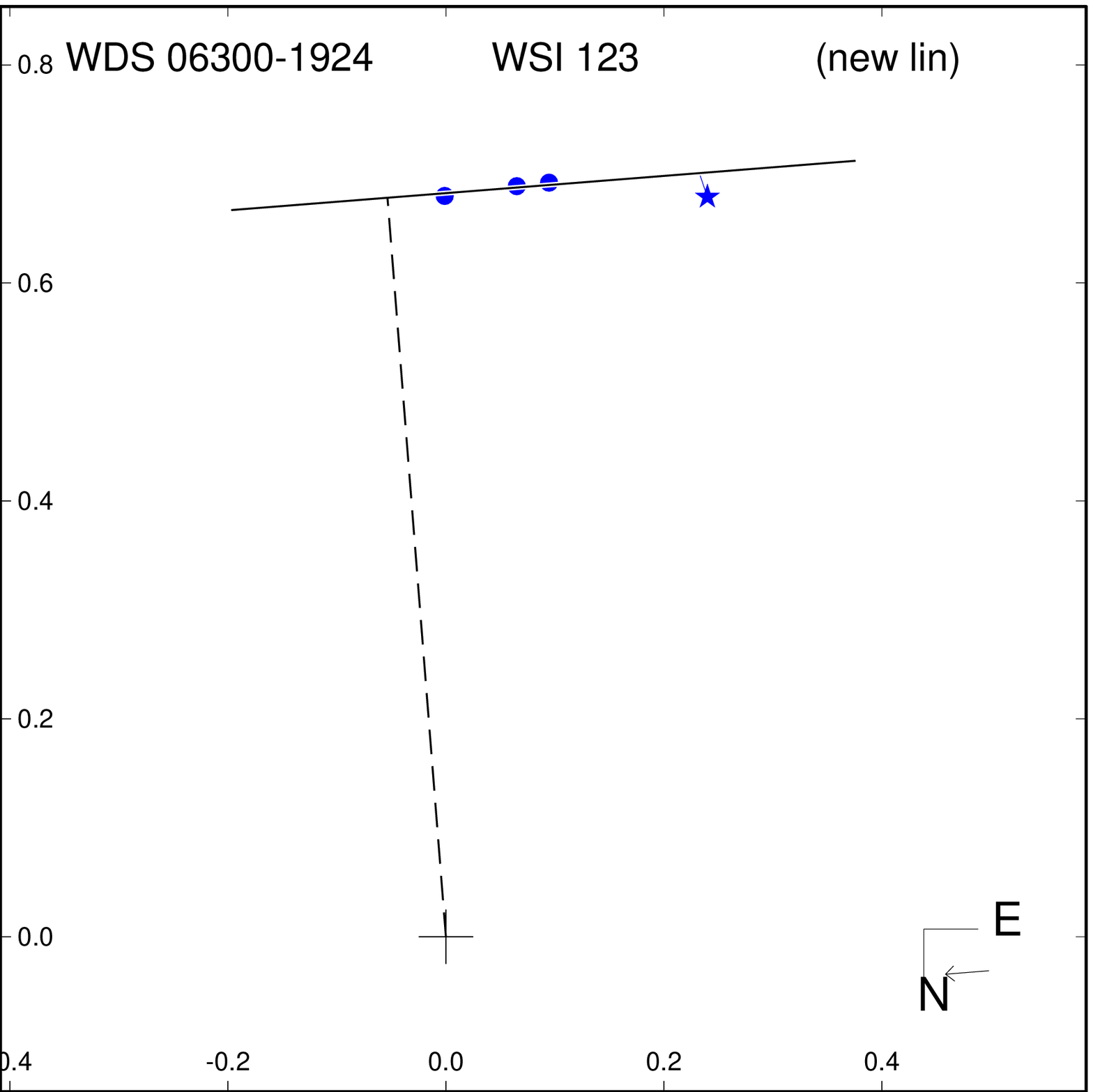} \epsfxsize 2.8in \epsffile{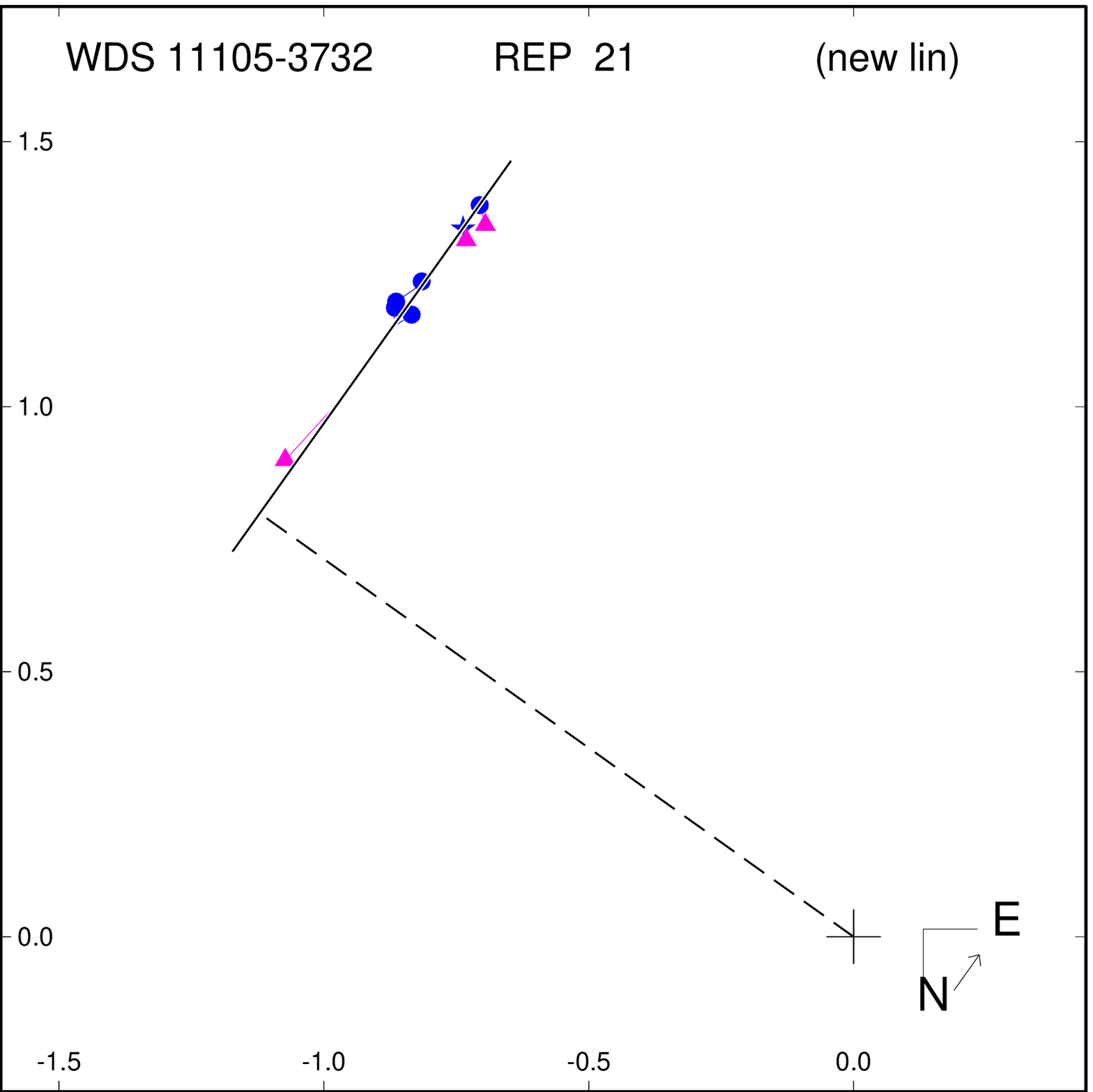}} 
\vskip 0.05in
{\epsfxsize 2.8in \epsffile{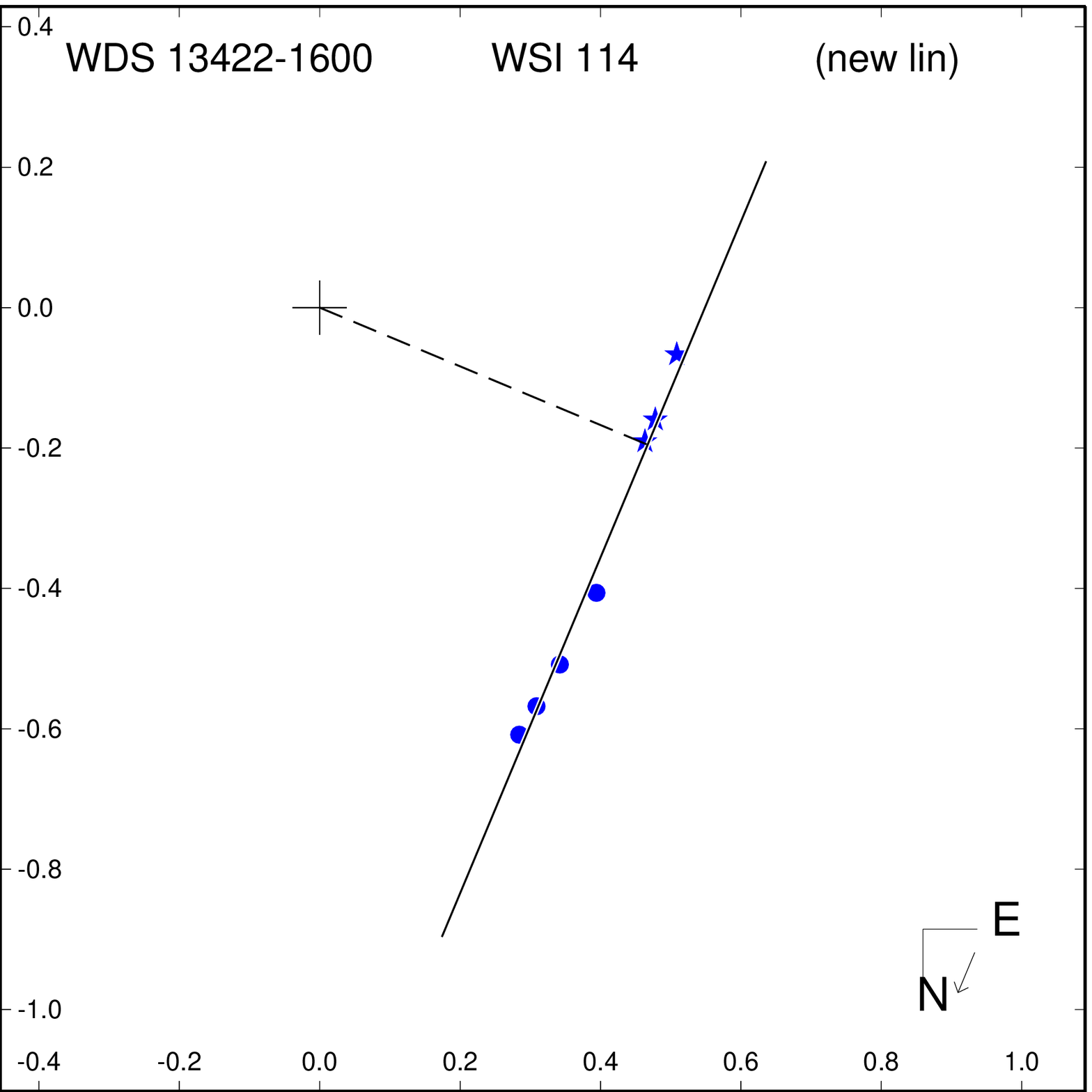} \epsfxsize 2.8in \epsffile{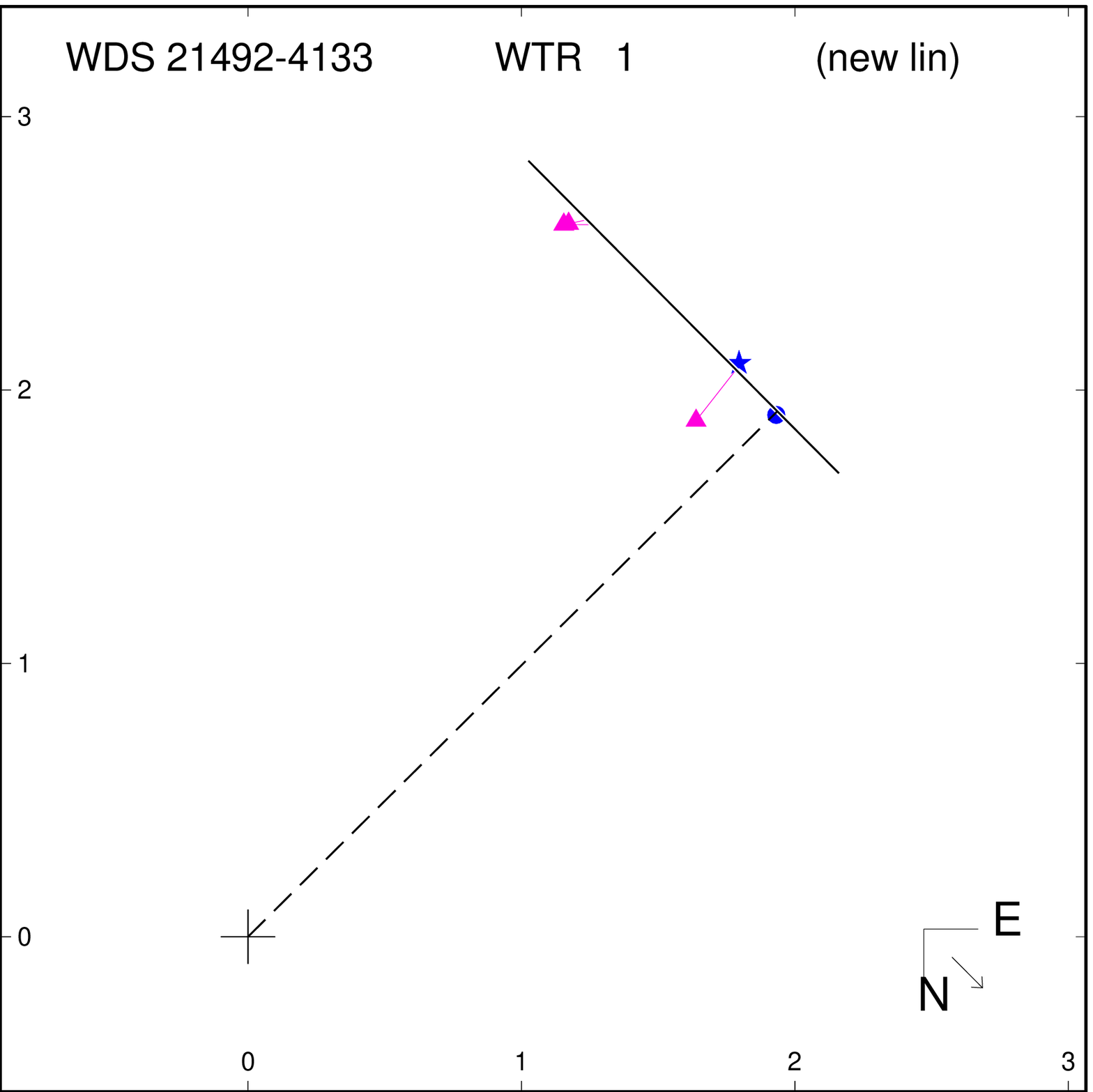}} 
\end{center}
\vskip -0.3in
\caption{\small New linear fits for the systems listed in Table 4 
and all data in the WDS database and Table 2. Symbols are the same as 
Figure 1. ``$O-C$" lines connect each measure to its predicted position along
the linear solution (shown as a thick solid line). An arrow in the lower 
right corner of each figure indicates the direction of motion. The scale, in 
arcseconds, is indicated on the left and bottom of each plot.}
\end{figure}

\vskip 0.1in

Table 5 gives ephemerides for each orbit or linear solution over the
years 2018 through 2023, in annual increments. Columns (1) and (2) 
are the same identifiers as in the previous tables, while columns 
(3+4), (5+6), ... (13+14) give predicted values of $\theta$ and 
$\rho$, respectively, for the years 2018.0,  2019.0, etc., through 
2023.0. All the orbit pairs are relatively fast moving, with mean 
motions of more than 6$^{\circ}$/yr. Notes to individual systems are
given in \S6.

\section{M dwarfs with no companion detected}

The selection of systems for this project was not blind and 
preference was given to systems previously known as double or having
parallax data from the CTIOPI program (Jao et al.\ 2005) that seemed
to indicate duplicity. Therefore, any duplicity rate we determine 
would be enriched and not representative of stars of this type. 
Despite this preselection, there were a large number of targets 
observed for which we did not detect a companion. 

%===========================================================================
Table 6 provides the complete list of unresolved red dwarfs obtained
on these observing runs. In some cases, known companions 
are not detectable due to the separation being wider than the field 
of view of 1\farcs8, or the magnitude difference being larger than 
detectable by the optical speckle camera. Due to the faintness of 
the primary targets, the companion must have $\Delta$m$~<~2$mag and 
$30mas~<~\rho~<~$1\farcs8. In this case, the upper limit is set by 
the minimum field of view when the object is centered for detection 
of unknown companions. As seen in Table 2, wider systems can be 
measured with {\it a priori} knowledge of the system or if they are 
seen while pointing the telescope. The usual procedure after moving 
the telescope to the approximate field was to step through larger 
fields of view obtained through 4$\times$4 or 2$\times$2 binning en 
route to a final un-binned field of about $6mas/pixel$. Data could be
taken in these binned fields to obtain measures of wider pairs. In 
some cases, pairs were too widely separated to be measured; often 
for these both components were observed separately. Finally, as some
of these targets are rather faint, an interference filter with a 
significantly larger FWHM ({\it Johnson V} as opposed to {\it Str\"{o}mgren y}) 
was used to allow enough photons to permit detection. However, use 
of this filter compromises the detection of the closest pairs. For 
these we set a lower separation limit of $50mas$. The cases where 
this filter was used are noted in Table 6. 

All individual observations, including a complete listing of each 
measure identifying the date of observation, resolution limit, 
filter and telescope, are given in the Catalog of Interferometric 
Measurements of Binary Stars\footnote{See Hartkopf et al.\ (2001b). 
The online version ({\tt http://ad.usno.navy.mil/wds/int4.html}) 
is updated frequently.}. Notes to individual systems reported here
are provided in \S6.

\section{Notes to Individual Systems}

{\bf WDS04073$-$2429 = BEU\phm{888}5 = LHS 1630}  (resolved, linear,
in WDS) : The proper motion (UCAC5; Zacharias et al.\ 2017) is 
$673.1mas/yr$, which seems to indicate the components are moving 
together with small changes in relative position, so the pair is 
classified as physical. However, their relative motion can be fit by
a line (see \S4.2, Table 4 and Figure 2). More data obtained over several 
years may determine if we have a companion which is optical, or 
if we happen to be catching the orbit on a long near-linear segment.

%UCAC4 (Zacharias et al.\ 2013) proper motion is 660.0 mas/yr.

{\bf WDS05000$-$0333 = JNN\phm{88}29 = SCR\phn J0459$-$0333} 
(unresolved, in WDS) : The companion has been measured multiple 
times, but only through red filters (Janson et al.\ 2012, 2014b). It
may be too faint in {\it Johnson V}. 

{\bf WDS05174$-$3522 = TSN\phm{888}1 = L\phn449-001} (unresolved, in
WDS) : The companion has only been measured with HST-FGS once at 
$47mas$ (Riedel et al.\ 2014), closer than our limit here with the 
{\it Johnson V} filter. This known pair is worth additional observations with 
large aperture high angular resolution techniques.

{\bf WDS06523$-$0510 = GJ\phn250} (resolved, in WDS) : The wide CPM 
pair, WNO\phn\phn17AB has many measures. Two 
unconfirmed companions to B have been measured. WSI\phn125Ba,Bb 
measured only in Table 2 and the much wider IR companion 
TNN\phm{888}6BC measured in Tanner et al.\ (2010). It is unknown if 
either of these are physical. We crudely estimate the $\Delta$m in V
as 0.5 for the Ba,Bb pair.

{\bf WDS07549-2920 = KUI\phm{88}32 = LHS\phn1955} (resolved, orbit, 
in WDS) : The first orbit of this pair. Based on these elements and 
the parallax ($74.36\pm1.13mas$; Winters et al.\ 2015), the 
resulting mass sum of 1.54$\pm$0.37\msun ~is suspiciously large. (see 
\S 4.1, Table 3 and Figure 1). It is possible that these preliminary 
orbital elements may aid future determinations and the planning of 
observing.

{\bf WDS08272$-$4459 = JOD\phm{888}5 = LHS\phn2010} (unresolved, in 
WDS) : The companion has only been measured once in the red ($914mas$
in 2008; Jodar et al.\ 2013). The companion is either too faint in 
the {\it Johnson V} observation or the companion is optical and has moved
to a separation too wide for detection. 

{\bf WDS08317$+$1924 = BEU\phm{88}12Aa,Ab = GJ\phm{8}2069} 
(unresolved, in WDS) : This pair of the multiple system has only 
been measured in the red or infrared. The companion is likely too 
faint in this {\it Johnson V} measurement for detection. The Ba,Bb pair is
resolved in Table 3. AB is CPM but is too wide for measurement here.

{\bf WDS10121$-$0241 = DEL\phm{888}3 = GJ\phm{88}381} (unresolved, 
in WDS) : The companion has only been measured in the red or 
infrared. It is likely too faint in this {\it Johnson V} measurement for 
detection.

{\bf WDS10430$-$0913 = WSI\phn112 = WT\phn1827} (resolved, in WDS) :
The companion is measured only in Table 2. It is unknown if it is 
physical. We crudely estimate the $\Delta$m in V as 1.7.

{\bf WDS11105$-$3732 = REP\phm{88}21 = TWA\phm{888}3}  (resolved, 
orbit or linear, in WDS) : The proper motion (UCAC4; Zacharias et 
al.\ 2013) is $107.3mas/yr$. While orbits with periods ranging from
236-800y have been determined, $``$the $\chi^2$ from the orbit fit
was indistinguishable from the linear fit" (Kellogg et al.\ 2017).
The solution presented in \S4.2, Table 4 and Figure 2 is a linear
fit to the data. Only time will tell if we have a companion which is
optical or we happen to be catching the orbit on a long near-linear 
segment.

{\bf 11354$-$3232 = GJ 433} (unresolved, not in WDS) : Detected as a
500d pair by Hipparcos (ESA 1997). However, according to Delfosse et
al.\ (2013), radial velocity coverage eliminates the Hipparcos 
result and the system just has one short-period planet.

{\bf WDS13422$-$1600 = WSI\phn114 = LHS 2783}  (resolved, linear, in
WDS) : Given the high proper motion of the PPMXL ($508.6mas/yr$; 
Roeser et al.\ 2010) and that from the CTIO-PI ($503.6mas/yr$;
Bartlett et al.\ 2017), it would indicate the stars are moving 
together. The measures can be fit by a line (see \S4.2, Table 4 and 
Figure 2), and thusfar do not seem to support the estimated period 
of 52y from Bartlett et al.\ (2017). However, based on this orbital 
period, the parallax, and an assumed total mass of 0.5\msun, a$''$ 
would be 0\farcs28, not too different from our measures (Table 2) of
about 0\farcs5. This tends to support the supposition that we are 
looking at a physical pair observed when the relative motion only 
appears to be linear. The pair should be monitored for variation 
from linearity.

%UCAC4 (Zacharias et al.\ 2013) proper motion is 508.6 mas/yr

{\bf WDS14540$+$2335 = REU\phm{888}2 = GJ 568} (resolved, orbit, in 
WDS) : The orbit of Heintz (1990) is improved here. Based on these 
elements and the parallax ($98.40\pm4.42mas$; van Leeuwen 2007) the
resulting mass sum is 0.261$\pm$0.083\msun. See \S 4.1, Table 3 and 
Figure 1.

{\bf 15301$-$0752 = G 152-31} (unresolved, not in WDS) : This 5.96y 
pair of Harrington \& Dahn (1988) should be resolvable (a$''$ = 
$496mas$ assuming $\Sigma~\cal{M}$ = 0.5\msun); therefore, it is 
assumed the $\Delta$m is higher than 2.5 and observation with a 
technique with a greater $\Delta$m sensitivity, such as adaptive 
optics, is appropriate.

{\bf WDS16240$+$4822 = HEN\phm{888}1Aa,Ab = GJ\phm{88}623} 
(unresolved, in WDS) : The companion has only been measured in the 
infrared or with HST-FGS. It likely has too large a $\Delta$m for V 
band detection here.

{\bf WDS17077$+$0722 = YSC\phm{88}62 = GJ 1210} (resolved, orbit, in
WDS) : This is the first orbit for this pair, whose first published 
measure (Horch et al.\ 2010) was made two years after that presented
in Table 2. Based on these elements and the parallax ($78.0\pm5.3mas$;
van Altena et al.\ 1995) the resulting mass sum is 
0.280$\pm$0.067\msun. ~See \S 4.1, Table 3 and Figure 1.  

{\bf WDS17119$-$0151 = LPM\phn629 = GJ 660} (resolved, orbit, in 
WDS) : The orbit of S\"{o}derhjelm (1999) is improved here. Based on
these elements and the parallax ($98.19\pm12.09mas$; van Leeuwen 
2007) the resulting mass sum is 0.40$\pm$0.16\msun. See \S 4.1, 
Table 3 and Figure 1.

{\bf WDS18387$-$1429 = HDS2641 = GJ\phn2138} (unresolved, in WDS) : 
The companion was measured by Hipparcos (ESA 1997) at $107mas$ and 
$\Delta$H$_p$ = 0.41. It would be expected to be resolved in our 
observation if near this location. Because it is not, the pair has 
either closed under $50mas$, was optical or was a false detection.

{\bf WDS19449$-$2338 = MTG\phm{888}4 = LP 869-26} (resolved, orbit, 
in WDS) : This is the first orbit for this pair. Based on these 
elements and the parallax ($67.87\pm1.1mas$; Bartlett et al.\ 2017)
the resulting mass sum is 0.283$\pm$0.086\msun. See \S 4.1, Table 3
and Figure 1.

{\bf 23018$-$0351 = GJ 886} (unresolved, not in WDS) : The 468.1d 
pair of Jancart et al.\ (2005) may have a separation close to our 
resolution limit, or slightly under it (a$''$ = $50mas$ assuming 
$\Sigma~\cal{M}$ = 0.5\msun). The $\Delta$m is unknown and may also 
be too high for our detection. This pair is worthy of additional 
observation.

\section{Conclusions}

In this paper, we report high resolution optical speckle 
observations of 336 M dwarfs that resulted in 113 resolved 
measurements of 80 systems and 256 other stars that gave no 
indication of duplicity within the detection limits of the 
telescope/system. We calculate orbits for six systems, two of which were 
revised and four which are first time orbits. All have short periods, 5-38y,
and these data may eventually assist in determining accurate masses.

\acknowledgements

%===========================================================================
The USNO speckle interferometry program has been supported by NASA 
and the SIM preparatory science program through NRA 98-OSS-007. This
research has made use of the SIMBAD database, operated at CDS, 
Strasbourg, France  and NASA's Astrophysics Data System. Thanks are also 
provided to the U.S.\ Naval Observatory for their continued support of the 
Double Star Program. The telescope operators and observing support personnel
of KPNO and CTIO continue to provide exceptional support for visiting 
astronomers. Thanks to Claudio Aguilero, Alberto Alvarez, Skip Andree, Bill 
Binkert, Gale Brehmer, Ed Eastburn, Angel Guerra, Hal Halbedal, Humberto 
Orrero, David Rojas, Hernan Tirado, Patricio Ugarte, Ricard Venegas, George
Will, and the rest of the KPNO and CTIO staff. Members of the RECONS team 
(JPS \& TJH) have been supported by NSF grants AST 05-07711, 09-08402 and 
14-12026. We would also like to thank Andrei Tokovinin for helpful comments.

\appendix

\section{Additional Measured Pairs}

Table A1 presents other, non-M dwarf pairs, observed during the runs
presented in Table 1. The first two columns identify the system by 
providing the WDS designation (based on epoch-2000 coordinates) and 
discovery designation. Columns three through five give the epoch of 
observation (expressed as a fractional Julian year), the position 
angle (in degrees), and the separation (in seconds of arc). The 
sixth column indicates the number of observations contained in the 
mean position. The last column is reserved for notes for these 
systems. 

\section{The Problem with WSI~138}

This pair was originally associated with LP 876-10. LP 876-10 was 
examined multiple times (Mamajek et al.\ 2013), none of which showed 
any hint of elongation. Tokovinin et al.\ (2015) also did not detect
it. Mamajek et al.\ effectively ruled this out an optical 
coincidence between the high proper motion LP 876-10 and a 
background star. The tentative conclusion is that a different pair 
was observed and the 2010 measure (see Table B1) was not of LP 
876-10, but instead of some other unidentified pair which may or may
not be a physical pair. While no nearby known pairs in the WDS
matches the approximate morphology of the pair, in this magnitude 
range an unknown double star would not be a surprise. Since we are 
unsure what star was examined the WDS does not provide a precise 
position, the magnitudes of the components are degraded, and it has 
been disassociated with Fomalhaut.

%\documentstyle[aj_pt4]{article}
%\begin{document}

\begin{deluxetable}{ccc}
\tablenum{1}
\tablewidth{0pt}
\tablecaption{Observing Runs}
\tablehead{
\colhead{Dates} & 
\colhead{Telescope} &
\colhead{Observers} \nl
}
\startdata
8-13 Nov 2005        & KPNO 4m & BDM \& WIH     \nl
9-13 Mar 2006        & CTIO 4m & BDM \& WIH     \nl
30 Jul - 10 Aug 2007 & KPNO 4m & BDM, WIH \& DR \nl
11-17 Jun 2008       & KPNO 4m & BDM \& DR      \nl
23-25 Jan 2010       & CTIO 4m & BDM \& JPS     \nl
1-3 Aug 2010         & CTIO 4m & WIH \& JPS     \nl
\enddata
\end{deluxetable}
%\end{document}

%\documentstyle[aj_pt4]{article}
%\begin{document}

%\scriptsize
\begin{deluxetable}{ll@{~}r@{~}lrrrrrrr}
\label{tab:linear}
\tablenum{4}
\tablewidth{0pt}
\tablecaption{New Linear Elements}
\tablehead{
\colhead{WDS} & 
\multicolumn{3}{c}{Discoverer} &
\colhead{x$_0$} & 
\colhead{a$_{\rm x}$} & 
\colhead{y$_0$} &
\colhead{a$_{\rm y}$} &
\colhead{T$_0$} &
\colhead{$\rho_0$} & 
\colhead{$\theta_0$} \nl
\colhead{$\alpha,\delta$ (2000)} & 
\multicolumn{3}{c}{Designation} & 
\colhead{($''$)} & 
\colhead{($''$/yr)} & 
\colhead{($''$)} &
\colhead{($''$/yr)} &
\colhead{(yr)} &
\colhead{($''$)} & 
\colhead{($\deg$)} \nl
\colhead{(1)} & 
\multicolumn{3}{c}{(2)} & 
\colhead{(3)} & 
\colhead{(4)} & 
\colhead{(5)} &
\colhead{(6)} &
\colhead{(7)} &
\colhead{(8)} & 
\colhead{(9)} \nl
}
\startdata
%                                          X0            XA             Y0            TA            T0       RHO0       THETA0
04073$-$2429 & BEU    &    5 &      &    0.250114 &    0.044070 &  $-$0.435759 &    0.025295 & 1994.338 &  0.502 &   29.85 \nl
05101$-$2341 & WSI    &  121 & Aa,Ab&    0.095884 & $-$0.022807 &     0.397173 &    0.005506 & 2023.416 &  0.409 &  166.43 \nl
06300$-$1924 & WSI    &  123 &      & $-$0.053621 & $-$0.028649 &     0.678105 & $-$0.002265 & 2020.057 &  0.680 &  184.52 \nl
11105$-$3732 & REP    &   21 &      & $-$1.120781 &    0.010277 &     0.798936 &    0.014417 & 1973.890 &  1.376 &  234.52 \nl
13422$-$1600 & WSI    &  114 &      &    0.467162 & $-$0.023533 &  $-$0.195311 & $-$0.056288 & 2010.631 &  0.506 &   67.31 \nl
21492$-$4133 & WTR    &    1 &      &    1.935622 &    0.042888 &     1.922322 & $-$0.044115 & 2012.841 &  2.665 &  136.81 \nl
\enddata
\end{deluxetable}%
%\end{document}

\vfill\eject


\begin{references}

%===========================================================================
\reference {} Bagnuolo, W.G., Jr., Mason, B.D., Barry, D.J.,
              Hartkopf, W.I.\ \& McAlister, H.A.\ 1992, AJ 103, 1399

\reference {} Balega, I.I., Balega, Yu.Yu.\ \& Malogolovets, E.V.\
              2010, Astrophysics Bulletin 65, 250

\reference {} Bartlett, J.L., Lurie, J.C., Riedel, A., Ianna, P.A., 
              Jao, W.-C., Henry, T.J., Winters, J.G., Finch, C.T.\ 
              \& Subasavage, J.P.\ \& 2017, AJ 154, 151

\reference {} Benedict, G.F., Henry, T.J., Franz, O.G., McArthur,
              B.E., Wasserman, L.H., Jao, W.-C., Cargile, P.A.,
              Dieterich, S.B., Bradley, A.J., Nelan, E.P.\ \&
              Whipple, A.L.\ 2016. AJ 152, 141

\reference {} Delfosse, X., Bonfils, X., Forveille, T., Udry, S.,
              Mayor, M., Bouchy, F., Gillon, M., Lovis, C., Neves,
              V., Pepe, F., Perrier, C., Queloz, D., Santos, N.C.\
              \& S\'{e}gransan, D.\ 2013, A\&A 553, 8

\reference {} Docobo, J.A., Balega, Y.Y., Ling, J.F., Tamazian, V.\
              \& Vasyuk, V.A.\ 2000, AJ 119, 2422

\reference {} Docobo, J.A., Tamazian, V.S., Balega, Y.Y., Andrade,
              M., Schertl, D., Weigelt, G., Campo, P.\ \& Palacios,
              M.\ 2008, A\&A 478, 187

\reference {} ESA 1997, The Hipparcos and Tycho Catalogues (ESA
              SP-1200) (Noordwijk: ESA)

\reference {} Forveille, T., Beuzit, J.-L., Delfosse, X., Segransan,
              D., Beck, F.\ \& Mayor, M.\ 1999 A\&A, 351, 619

\reference {} Harrington, R.S.\ \& Dahn, C.C.\ 1988, AJ 96, 718

\reference {} Hartkopf, W.I.\ \& Mason, B.D.\ 2014, IAU DS Circular
              \#184, 1

\reference {} Hartkopf, W.I.\ \& Mason, B.D.\ 2015, $``$Catalog of 
              Rectilinear Elements"\footnote{see current version at 
              {\tt http://ad.usno.navy.mil/wds/lin1.html}.}

\reference {} Hartkopf, W.I., Mason, B.D.\ \& McAlister, H.A.\ 1996,
              AJ 111, 370

\reference {} Hartkopf, W.I., Mason, B.D.\ \& Worley, C.E.\ 2001a,
              AJ 122, 3472\footnote{see current version at 
              {\tt http://ad.usno.navy.mil/wds/orb6.html}.}

\reference {} Hartkopf, W.I., McAlister, H.A.\ \& Franz, O.G.\ 1989,
              AJ 98, 1014

\reference {} Hartkopf, W.I., McAlister, H.A., \& Mason, B.D. 2001b,
              AJ 122, 3480\footnote{see current version at 
              http://ad.usno.navy.mil/wds/int4.html.}

\reference {} Hartkopf, W.I., Tokovinin, A.\ \& Mason, B.D.\ 2012,
              AJ 143, 42

\reference {} Heintz, W.D.\ 1986, A\&AS 65, 411

\reference {} Heintz, W.D.\ 1990, A\&AS 82, 65

%\reference {} Heintz, W.D.\ \& Cantor, B.\ 1992, Obs 112, 286

\reference {} Henry, T.J., Franz, O.G., Wasserman, L.H., Benedict
              G.F., Shelus, P.J., Ianna, P.A., Kirkpatrick, J.D.\ \&
              McCarthy, D.W., Jr.\ 1999, ApJ 512, 864

\reference {} Henry, T.J., Jao, W.-C., Subasavage. J.P., Beaulieu,
              T.D., Ianna, P.A., Costa, E.\ \& M\'{e}ndez, R.A.\
              2006, AJ 132, 2360

%\reference {} Henry, T.J., Walkowicz, L.M., Barto, T.C.\ \&
%              Golimowski, D.A.\ 2002, AJ 123, 2002

%\reference {} Herschel, W.\ 1803, Phil.\ Trans.\ R.\ Soc.\ 93, 339

\reference {} Horch, E.P., Bahi, L.A.P., Gaulin, J.R., Howell, S.B.,
              Sherry, W.H., Galle, R.B.\ \& van Altena, W.F.\ 2012, 
              AJ 143, 10

\reference {} Horch, E.P., Falta, D., Anderson, L.M., DeSousa, M.D.,
              Miniter, C.M., Ahmed, T.\ \& van Altena, W.F.\ 2010,
              AJ 139, 205

\reference {} Horch, E.P., Gomez, S.C., Sherry, W.H., Howell, S.B.,
              Ciardi, D.R., Anderson, L.M.\ \& van Altena, W.F.\
              2011, AJ 141, 45

\reference {} Jancart, S., Jorrisen, A., Babusiaux, C., \& Pourbaix,
              D. 2005, A\&A 442, 365

\reference {} Janson, M., Bergfors, C., Brandner, W., Bonnefoy, M.,
              Schlieder, J., K\"{o}hler, R., Hormuth, F., Henning,
              T.\ \& Hippler, S. 2014b, ApJS 214, 17

\reference {} Janson, M., Bergfors, C., Brandner, W., Kudryavtseva,
              N., Hormuth, F., Hippler, S.\ \& Henning, T.\ 2014a, 
              ApJ 789, 102

\reference {} Janson, M., Hormuth, F., Bergfors, C., Brandner, W.,
              Hippler, S., Daemgen, S., Kudryavtseva, N., Schmalzl,
              E., Schnupp, C.\ \& Henning, T. 2012, ApJ 754, 44

%\reference {} Jao, W.-C., Henry, T.J., Subasavage, J.P., Bean, J.L.,
%              Costa, E., Ianna, P.A.\ \& Mendez, R.A.\ 2003, AJ 125,
%              332

\reference {} Jao, W.-C., Henry, T.J., Subasavage, J.P., Brown,
              M.A., Ianna, P.A., Bartlett, J.L., Costa, E.\ \&
              Mendez, R.A.\ 2005, AJ 129, 1954

\reference {} Jodar, E., Perez-Garrido, A., Diaz-Sanchez, A., Villo,
              I., Rebolo, R.\ \& Perez-Prieto, J.A.\ 2013, MNRAS
              429, 859

\reference {} Kellogg, K., Prato, L., Torres, G., Schaefer, G.H., 
              Avilez, I., Ruiz-Rodr\'{i}quez, D., Wasserman, L.H.,
              Bonanos, A.Z., Guenther, E.W., Neuh\"{a}user, R., 
              Levine, S.E., Bosh, A.S., Morzinski, K.M., Close, L.,
              Bailey, V., Hinz, P.\ \& Males, J.R.\ 2017, ApJ 844, 
              168

\reference {} Kervalla, P., Merand, A., Ledoux, C., Demory, B.-O.\ 
              \& Le Bouquin, J.-B.\ 2016, A\&A 593, 127

\reference {} K\"{o}hler, R., Ratzka, T.\ \& Leinert, Ch.\ 2012 A\&A
              541, 29

\reference {} Mamajek, E.E., Bartlett, J.L., Seifahrt, A., Henry,
              T.J., Dieterich, S.B., Lurie, J.C., Kenworthy, M.A.,
              Jao, W.-C., Riedel, A.R., Subasavage, J.P., Winters,
              J.G., Finch, C.T., Ianna, P.A.\ \& Bean, J.\ 2013, AJ
              146, 154

\reference {} Mason, B.D.\ \& Hartkopf, W.I.\ 2012, IAU DS Circular
              \#178, 1

\reference {} Mason, B.D., Hartkopf, W.I., Gies, D.R., Henry, T.J.\
              \& Helsel, J.W.\ 2009, AJ 137, 3358

\reference {} Mason, B.D., Hartkopf, W.I., Raghavan, D., Subasavage,
              J.P., Roberts, Jr., L.C., Turner, N.H.\ \& ten
              Brummelaar, T.A.\ 2011, AJ 142, 176

\reference {} Mason, B.D., Wycoff, G.L., Hartkopf, W.I., Douglass,
              G.G.\ \& Worley, C.E.\ 2001, AJ 122,
              3466\footnote{see current version at 
              {\tt http://ad.usno.navy.mil/wds/}.}

\reference {} McAlister, H.A., Hartkopf, W.I., Hutter, D.J.\ \& Franz,
              O.G.\ 1987, AJ 93, 688

\reference {} Miles, K.N.\ \& Mason, B.D.\ 2016, IAU DS Circular
              \#190, 1

\reference {} Montagnier, G., S\'{e}gransan, D., Beuzit, J.\L.,
              Forveille, T., Delorme, P., Delfosse, X., Perrier, C.,
              Udry, S., Mayor, M., Chauvin, G., Lagrange, A.-M.,
              Mouillet, D., Fusco, T., Gigan, P.\ \& Stadler, E.\
              2006, A\&AL 460, 19

\reference {} Prieur J.-L., Scardia, M., Pansecchi, L., Argyle,
              R.W., Zannuta, A.\ \& Aristidi, E.\ 2014, AN 335, 817

%\reference {} Reid, I.N., Kilkenny, D.\ \& Cruz, K.L.\ 2002, AJ 123,
%              2822

\reference {} Riedel, A.R., Finch, C.T., Henry, T.J., Subasavage,
              J.P., Jao, W.-C., Malo, L., Rodriguez, D.R., White,
              R.J., Gies, D.R., Dieterich, S.B., Winters, J.G.,
              Davison, C.L., Nelan, E.P., Blunt, S.C., Cruz, K.L.,
              Rice, E.L.\ \& Ianna, P.A.\ 2014, AJ147, 85

\reference {} Roeser, S., Demleitner, M.\ \& Schilbach, E.\ 2010, AJ
              139, 2440

\reference {} Schulz, A.B., Hart, H.M., Hershey, J.L., Hamilton, 
              F.C., Kochte, M., Bruhweiler, F.C., Benedict, G.F.,
              Caldwell, J., Cunningham, C., Franz, O.G., Keyes,
              C.D.\ \& Brandt, J.C.\ 1998, PASP 110, 31

\reference {} S\'{e}gransan, D., Delfosse, X., Forveille, T.,
              Beuzit, J.-L., Udry, S., Perrier, C., \& Mayor, M.\
              2000, A\&A 364, 665

\reference {} S\"{o}derhjelm, S.\ 1999 A\&A 341, 121

\reference {} Tamazian, V.S., Docobo, J.A., Melikian, N.D., Balega,
              Y.Y.\ \& Karaperian, A.A.\ 2005, AJ 130, 2847

\reference {} Tanner, A.M., Gelino, C.R.\ \& Law, N.M.\ 2010, PASP
              122, 1195

%\reference {} Tokovinin, A.\ 2012, AJ 144, 56

\reference {} Tokovinin, A., Mason, B.D.\ \& Hartkopf, W.I.\ 2010,
              AJ 139, 1105

\reference {} Tokovinin, A., Mason, B.D.\ \& Hartkopf, W.I.\ 2014,
              AJ 147, 123

\reference {} Tokovinin, A., Mason, B.D., Hartkopf, W.I., Mendez,
              R.A.\ \& Horch, E.P.\ 2015, AJ 150, 50

\reference {} Tokovinin, A., Mason, B.D., Hartkopf, W.I., Mendez,
              R.A.\ \& Horch, E.P.\ 2016, AJ 151, 153

\reference {} Tokovinin, A., Mason, B.D., Hartkopf, W.I., Mendez,
              R.A.\ \& Horch, E.P.\ 2018, AJ ({\it submitted})

\reference {} Tokovinin, A., Mason, B.D., Hartkopf, W.I., Mendez,
              R.A.\ \& Horch, E.P.\ 2018 ({\it in preparation})

\reference {} van Altena, W.F., Lee, J.T.\ \& Hoffleit, E.D. 1995,
              {\it The General Catalogue of Trigonometric (stellar)
              Parallaxes} (4th ed.; New Haven, CT: Yale Univ.\
              Observatory)

\reference {} van Leeuwen, F.\ 2007, A\&A 474, 653

\reference {} Ward-Duong, K. Patience, J., De Rosa, R.J., Bulger,
              J., Rajan, A., Goodwin, S.P., Parker, R.J., McCarthy,
              D.W.\ \& Kulesa, C.\ 2015, MNRAS 449, 2618

%\reference {} Winters, J.G.\ 2015, PhD thesis, Georgia State Univ.\

\reference {} Winters, J.G., Henry, T.J., Jao, W.-C., Subasavage,
              J.P., Finch, C.T.\ \& Hambly, N.C.\ 2011, AJ 141, 21

\reference {} Winters, J.G., Henry, T., Lurie, J., Hambly, N., Jao,
              W.-C., Bartlett, J., Boyd, M., Bieterich, S., Finch, 
              C., Hosey, A., Ianna, P., Riedel, A., Slatten, K.\ \& 
              Subasavage, J.\ 2015, AJ 149, 5

\reference {} Winters, J.G., Sevrinsky, R.A., Jao, W.-C., Henry,
              T.J., Riedel, A.R., Subasavage, J.P., Lurie, J.C.,
              Finch, C.T.\ \& Ianna, P.A.\ 2017, AJ 153, 14

\reference {} Zacharias, N., Finch, C.\ \& Frouard, J.\ 2017, AJ
              153, 166

\reference {} Zacharias, N., Finch, C.T., Girard, T.M., Henden, A.,
              Bartlett, J.L., Monet, D.G.\ \& Zacharias, M.I. 2013,
              AJ 145, 44

\reference {} Zirm, H.\ 2003, IAU DS Circular \#151, 1

\end{references}
\end{document}